\documentclass[12pt]{article}
\usepackage{color}
\usepackage{graphicx}
\usepackage{slashed}
\usepackage{amsmath,amstext,amsgen,amsbsy,amsopn,amsfonts,theorem}

\usepackage{geometry}
\begin{document}
\title{Radiative decay widths of ground and excited states of vector charmonium and bottomonium}
\author{Hluf Negash$^{a}$ and Shashank Bhatnagar$^{b}$}
\maketitle
$^{a}$Department of Physics, Samara University, Samara, Ethiopia\\
$^{b}$Department of Physics, University Institute of Sciences, Chandigarh University, Mohali-140413, India\\

\textbf{Abstract}\\
In this work we study the radiative decay widths of vector
quarkonia for the process of
$J/\psi(nS)\rightarrow\eta_{c}(nS)\gamma$ and
$\Upsilon(nS)\rightarrow\eta_{b}(nS)\gamma$ (for principal quantum
numbers $n=1, 2, 3$) in the framework of Bethe-Salpeter equation
under the covariant instantaneous ansatz using a $4\times 4$ form
of BSE. The parameters of the framework were determined by a fit
to the mass spectrum of ground states of pseudoscalar and vector
quarkonia, such as; $\eta_{c}$, $\eta_{b}$, $J/\psi$ and
$\Upsilon$. These input parameters so fixed were found to give
good agreements with data on mass spectra of ground and excited
states of pseudoscalar and vector quarkonia, leptonic decay
constants of pseudoscalar and vector quarkonia, two photon decays
and two gluon decays of pseudoscalar quarkonia in our recent
paper. With these input parameters so fixed, the radiative decay
widths of ground (1S) and excited (2S, 3S) states of heavy vector
quarkonia ($J/\Psi$ and $\Upsilon$) are calculated and found to be
in reasonable agreement with data.

\section{Introduction}
Studies on mass spectra and decays of heavy quarkonia
($c\overline{c}$, and $b\overline{b}$) have become a hot topic in
recent years, due to observation of many new states at various high
energy accelerators at BABAR, Belle, CLEO and BES-III collaborations
\cite{babar09,belle10,cleo01,olive14}. All this has opened up new
challenges in theoretical understanding of heavy hadrons and provide
an important tool for exploring the structure of these simplest
bound states in QCD and for studying the non-perturbative (long
distance) behavior of strong interactions.

As regards the dynamical framework, to investigate these properties
is concerned, many non-perturbative approaches, such as Lattice QCD
\cite{mcnielle12}, Chiral perturbation theory \cite{gasser84}, QCD
sum rules \cite{veli12}, heavy quark effective theory
\cite{neubert94}, N.R.QCD \cite{bodwin}, dynamical-equation based
approaches like Schwinger-Dyson equation and Bethe-Salpeter equation
(BSE) \cite{smith69,mitra01,alkofer01,wang10,koll,bhatnagar92}, and
potential models \cite{smriti16} have been proposed to deal with the
long distance property of QCD.
\bigskip

Bethe-Salpeter equation (BSE)
\cite{mitra01,bhatnagar92,mitra99,bhatnagar91,bhatnagar14} is a
conventional approach in dealing with relativistic bound state
problems. From the solutions, we can obtain useful information about
the inner structure of hadrons, which is also crucial in treating
hadronic decays. The BSE framework which is firmly rooted in field
theory, provides a realistic description for analyzing hadrons as
composite objects. Despite its drawback of having to input
model-dependent kernel, these studies have become an interesting
topic in recent years, since calculations have shown that BSE
framework using phenomenological potentials can give satisfactory
results as more and more data are being accumulated. The BSE is
frequently adopted as starting point of QCD inspired models, due to
the fact that this equation has a firm base in quantum field theory.

\bigskip
In a recent work \cite{hluf15,hluf16}, we employed a $4\times 4$
representation for two-body ($q\overline{q}$) BS amplitude for
calculating both the mass spectra as well as the transition
amplitudes for various processes. However, the price one to pay in
this approach is to solve a coupled set of Salpeter equations for
both pseudoscalar and vector quarkonia. However, in \cite{hluf16},
we explicitly showed that these coupled Salpeter equations can
indeed get decoupled in the heavy-quark approximation, leading to
mass spectral equations with analytical solutions for both masses,
as well as eigenfunctions for all the ground and excited states of
pseudoscalar and vector $c\overline{c}$ and $b\overline{b}$ systems
in an approximate harmonic oscillator basis. These analytical forms
of eigen functions for ground and excited states so obtained were
used to evaluate the transition amplitudes for different processes
in \cite{hluf16}. Thus in \cite{hluf16}, we had calculated the mass
spectrum, weak decay constants, two photon decay widths and two
gluon decay widths of ground (1S) and radially excited (2S, 3S,...)
states of pseudoscalar charmoniuum and bottomonium such as
$\eta_{c}$ and $\eta_{b}$, as well as the mass spectrum and leptonic
decay constants of ground state (1S), excited (2S, 1D, 3S, 2D, 4S
and 3D) states of vector charmonium and bottomonium such as $J/\psi$
and $\Upsilon$, using this formulation of $4\times 4$ Bethe-Salpeter
equation under covariant Instantaneous Ansatz (CIA). Our results
were in good agreement with data (where ever available) and other
models. However, in all the above processes, the quark anti-quark
loop involved a single hadron-quark vertex, which was simple to
handle.

However for the transitions such as $V\rightarrow P+\gamma$, the
process requires calculation of triangle quark-loop diagram
involving two hadron-quark vertices and is difficult to evaluate in
BSE-CIA, which give rise to complexities in amplitudes. However in
\cite{wang5,chen}, they demonstrated an explicit mathematical
procedure for handling such problems in $4\times4$ representation of
BSE. Thus, in the present work, we will precisely apply an
instantaneous formalism employing $4\times4$ BSE under CIA for
transitions involving the process, $V\rightarrow P\gamma$, where
such problems do not enter in the calculations of \cite{hluf16}.

\bigskip
This paper is organized as follows. In Sec. \textbf{2}, we give the
formulation of $4\times4$ BSE under CIA. In Sec. \textbf{3}, we give
the derivation of the hadronic process $V\rightarrow P+\gamma$ in
the framework of $4\times 4$ BSE under CIA and calculate its
radiative decay widths. The numerical results for radiative decay
widths of the processes are worked out. Section \textbf{4} is
reserved for discussion and conclusions.
\bigskip

\section{Formulation of BSE under CIA}

We give a short derivation of Salpeter equations in this section,
giving only the main steps. The 4D BSE for $q\bar{q}$ comprising of
equal mass fermionic quarks of momenta $p_{1,2}$, and masses
$m_{1}=m_{2} (=m)$ respectively is written in $4\times 4$
representation as:

\begin{equation}
\ (\slashed{p}_{1}-m_{1})\Psi(P,q)(\slashed{p}_{2}+m_{2}) =
\frac{i}{(2\pi)^{4}}\int d^{4}q'K(q,q')\Psi(P,q')
\end{equation}

where the $4\times 4$ BS wave function is sandwiched between the
inverse propagators of the quark and the ant-quark, whose individual
momenta $p_{1,2}$ are related to the internal momentum $q$ and total
momentum $P$ of hadron of mass $M$ as, $p_{1,2\mu} =
\frac{1}{2}P_{\mu} \pm q_{\mu}$. We further decompose the internal
momentum, $q_\mu$ as the sum of its transverse component,
$\hat{q}_\mu=q_\mu-(q\cdot P)P_\mu/P^2$ (which is orthogonal to
total hadron momentum $P_\mu$), and the longitudinal component,
$\sigma P_\mu = (q\cdot P)P_\mu/P^2$, (which is parallel to
$P_\mu$). Thus, $q_\mu=(M\sigma,\widehat{q})$, where the transverse
component, $\widehat{q}$ is an effective 3D vector, while the
longitudinal component, $M\sigma$ plays the role of the time
component. The 4-D volume element in this decomposition is,
$d^4q=d^3\hat{q}Md\sigma$. To obtain the 3D BSE and the hadron-quark
vertex, use an Ansatz on the BS kernel $K$ in Eq. (1) which is
assumed to depend on the 3D variables $\hat{q}_\mu$,
$\hat{q}_\mu^\prime$ as,
\begin{equation}
 K(q,q') = K(\hat{q},\hat{q}')
\end{equation}
Hence, the longitudinal component, $M\sigma$ of $q_{\mu}$, does not
appear in the form $K(\hat{q},\hat{q}')$ of the kernel and we define
3D wave function $\psi(\hat{q})$ as:
\begin{equation}
\psi(\hat{q}) = \frac{i}{2\pi}\int Md\sigma\Psi(P,q)
\end{equation}
Substituting Eq.(3) in eq.(1), with definition of kernel in eq.(2),
we get a covariant version of Salpeter equation,
\begin{equation}
\ (\slashed{p}_{1}-m_{1})\Psi(P,q)(\slashed{p}_{2}+m_{2}) = \int
\frac{d^{3}\hat{q}'}{(2\pi)^{3}} K(\hat{q},\hat{q}')\psi(\hat{q}'),
\end{equation}

and the 4D BS wave function can be written as,
\begin{equation}
\Psi(P,q) = S_{F}(p_{1})\Gamma(\hat{q})S_{F}(-p_{2})
\end{equation}
where
\begin{equation}
\Gamma(\hat{q})=\int \frac{d^{3}\hat{q}'}{(2\pi)^{3}}
K(\hat{q},\hat{q}')\psi(\hat{q}')
\end{equation}
 plays the role of hadron-quark vertex function. Following a
 sequence of steps given in \cite{hluf15,hluf16}
we obtain four Salpeter equations:
\begin{eqnarray}
 &&\nonumber(M-2\omega)\psi^{++}(\hat{q})=-\Lambda_{1}^{+}(\hat{q})\Gamma(\hat{q})\Lambda_{2}^{+}(\hat{q})\\&&
   \nonumber(M+2\omega)\psi^{--}(\hat{q})=\Lambda_{1}^{-}(\hat{q})\Gamma(\hat{q})\Lambda_{2}^{-}(\hat{q})\\&&
 \psi^{+-}(\hat{q})=\psi^{-+}(\hat{q})=0
\end{eqnarray}
with  the energy projection operators,
$\Lambda_{j}^{\pm}(\hat{q})=\frac{1}{2\omega_{j}}\bigg[\frac{\slashed{P}\omega_{j}}{M}\pm
I(j)(im_{j}+\slashed{\hat{q}})\bigg]$,
$\omega_{j}^{2}=m_{j}^{2}+\hat{q}^{2}$, and $I(j)=(-1)^{j+1}$ where
$j=1, 2$ for quarks and anti-quarks respectively. The projected wave
functions, $\psi^{\pm\pm}(\hat{q})$ in Salpeter equations are
obtained by the operation of the above projection operators on
$\psi(\widehat{q})$ (for details see \cite{hluf15,hluf16} as,
\begin{equation}
 \psi^{\pm\pm}(\hat{q})=\Lambda_{1}^{\pm}(\hat{q})\frac{\slashed{P}}{M}\psi(\hat{q})
 \frac{\slashed{P}}{M}\Lambda_{2}^{\pm}(\hat{q}).
\end{equation}

To obtain the mass spectral equation, we have to start with the
above four Salpeter equations and solve the instantaneous Bethe
Salpeter equation. However, the last two equations do not contain
eigenvalue $M$, and are thus employed to obtain constraint
conditions on the Bethe - Salpeter amplitudes associated with
various Dirac structures in $\psi(\widehat{q})$, as shown in details
in \cite{hluf16}. The framework is quite general so far. In fact the
above four equations constitute an eigenvalue problem that should
lead to evaluation of mass spectra of pseudoscalar and vector
charmonium and bottomonium states such as $\eta_{c}$, $\eta_{b}$,
$J/\psi$, and $\Upsilon$ (see \cite{hluf16}). The numerical
results\cite{hluf16} of mass spectra of P and V quarkonia
participating in the radiative decays studied in this paper are
listed in Section 3. We now give details of calculation of decay
widths for the process, $V\rightarrow P+\gamma$ in the next section.

\section{Electromagnetic transition of $V\rightarrow P+\gamma$}
The lowest order, Feynman diagrams for the process, $V \rightarrow P
\gamma$, are given in Fig.1, where $V (1^{--})$ and $P (0^{-+})$ are
vector and pseudoscalar quarkonia respectively. The second diagram
is obtained from the first one by reversing the directions of
internal fermionic lines.
\begin{figure}[h]
\centering
    \includegraphics[width=14cm,]{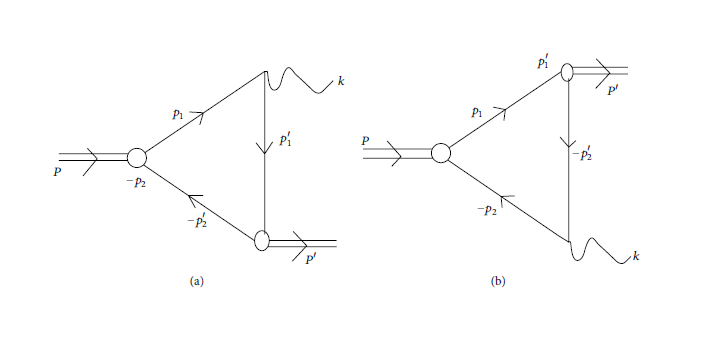}\\
       \caption{Feynman diagram for the first and second figures corresponding to the quark and anti-quark emitting
the photon respectively, for the transition $V\rightarrow
P\gamma$.}\label{4}
\end{figure}
The electromagnetic transition amplitude $M_{fi}$, corresponding to
Fig.1a, and 1b is given by the one-loop momentum integral as in Ref.
\cite{chen,mitra,bhatnagar12}:
\begin{equation}
M_{fi}=-ie_{q}\int\frac{d^{4}q}{(2\pi)^{4}}Tr[\overline{\Psi}_{P}(P',q')\slashed{\epsilon}\Psi_{V}(P,q)S^{-1}_{F}(-p_{2})
 +\overline{\Psi}_{P}(P',q')S^{-1}_{F}(p_{1})\Psi_{V}(P,q)\slashed{\epsilon}]
\end{equation}
where $\Psi_{P}$ and $\Psi_{V}$ are the 4D BS wave functions of
pseudoscalar and vector quarkonia involved in the process, and are
expressed as,  $\Psi_{V}(P,q) =
S_{F}(p_{1})\Gamma(\hat{q})S_{F}(-p_{2})$ and
$\overline{\Psi}_{P}(P',q')=
S_{F}(-p'_{2})\Gamma(\hat{q'})S_{F}(p'_{1})$. $e_{q}$ is the
electric charge of quarks. $\epsilon_{\mu}$ is the polarization
vector of the emitted photon. We have used the momentum relations:
$P=p_{1}+p_{2}$, $P'=p'_{1}+p'_{2}$, $k=P-P'$, and $p_{2}=p'_{2}$
for first diagram. However for second diagram, the last momentum
relation is $p_{1}=p'_{1}$, while the other relations are the same
as in first diagram. Here, the second term (coming from Fig.1b) in
Eq. (9) gives the same contribution as the first term (coming from
Fig.1a), and hence we can write $M_{fi}$ as:
\begin{equation}
M_{fi}=-2ie_{q}\int\frac{d^{4}q}{(2\pi)^{4}}Tr[\overline{\Psi}_{P}(P',q')\slashed{\epsilon}\Psi_{V}(P,q)S^{-1}_{F}(-p_{2})]
\end{equation}
with an overall multiplicative factor of 2.

This equation can be reduced to 3D forms by defining the 3D wave
function, $\psi(\hat{q}) = \frac{i}{2\pi }\int Md\sigma\Psi(P,q)$.
Thus, one can write the instantaneous Bethe-Salpeter form for the
electromagnetic transition amplitude between two bound states as
(see Ref. \cite{wang1}):
\begin{equation}
M_{fi}=-\frac{2e_{q}}{M}\int\frac{d^{3}\hat{q}}{(2\pi)^{3}}Tr\{\slashed{P}\overline{\psi}_{P}^{++}(\hat{q}')
\slashed{\epsilon}\psi_{V}^{++}(\hat{q})-\slashed{P}\overline{\psi}_{P}^{--}(\hat{q}')
\slashed{\epsilon}\psi_{V}^{--}(\hat{q})\}
\end{equation}
where, we resolve the amplitude corresponding to the first term
(i.e. Fig.1a) in the previous equation into $++$, and $--$
components, where the dominant contribution comes from the $++$
components, while the contribution from the $--$ components is less
than $1$ percent \cite{wang3}. Thus we can to a good approximation,
write, the total amplitude for the process $V\rightarrow P\gamma$ in
terms of only the $++$ components as:
\begin{equation}
M_{fi}=-\frac{2e_{q}}{M}\int\frac{d^{3}\hat{q}}{(2\pi)^{3}}Tr\{\slashed{P}\overline{\psi}_{P}^{++}(\hat{q}')
\slashed{\epsilon}\psi_{V}^{++}(\hat{q})\}
\end{equation}
where, $\hat{q}'=\hat{q}+\frac{1}{2}\hat{P}'$ and $M$ is the mass of
the initial quarkonia. The relativistic BS wave function of
$P(0^{-+})$ meson bound state is taken as in Ref.\cite{hluf16}:
\begin{equation}
\psi_{P}(\hat{q}') =N'_{P}
[M'+\slashed{P'}+\frac{\slashed{\hat{q}'}\slashed{P}'}{m}]\gamma_{5}\phi_{P}(\hat{q}')
\end{equation}
where $P'$ is the momentum of the final state P-quarkonia, $N'_{P}$
is the BS normalization of the final state P-quarkonia and $M'$ is
the mass of the final state P-quarkonia.

The relativistic BS wave function of $V (1^{--})$ quarkonium can
ultimately be written as in Ref.\cite{hluf16}:
\begin{equation}
\psi_{V}(\hat{q})=N_{V}[M\slashed{\varepsilon}+\hat{q}.\varepsilon\frac{M}{m}+\slashed{\varepsilon}\slashed{P}+
\frac{\slashed{P}\hat{q}.\varepsilon}{m}-\frac{\slashed{P}\slashed{\varepsilon
}\slashed{\hat{q}}}{m}]\phi_{V}(\hat{q})
\end{equation}

Here, $\varepsilon$ is the polarization vector of the vector
quarkonia and $N_{V}$ is the BS normalizer of the initial state
quarkonia. We wish to mention that Eq.(13) above is obtained by
starting from the most general expression for $\Psi(P,q)$ in Eq.(17)
of \cite{hluf16}. The general decomposition of instantaneous wave
function in the center of mass frame of dimensionality $M$ is given
in Eq.(18) of \cite{hluf16}. This instantaneous wave function
$\Psi^P(\widehat{q})$ is put into the last two Salpeter equations,
and the constraint conditions on the amplitudes (given in Eq.(19) of
\cite{hluf16}) are obtained. We thus obtain Eq.(20) of
\cite{hluf16}. This is then put into the first two Salpeter
equations to obtain the two coupled equations (Eq.(21) of
\cite{hluf16}). Solving them simultaneously leads to Eq.(13) of
present paper (which is Eq.(23) in \cite{hluf16}), written in terms
of $\phi_P$. Similarly we obtain $\Psi_V(\widehat{q})$ in Eq.(14)
written in terms of $\phi_V$. These scalar functions
$\phi_{P,V}(\widehat{q})$ satisfy the harmonic oscillator equation,
Eq.(37) of \cite{hluf16}, whose solutions are the worked out by
using the power series method in Eq.(37-41) of \cite{hluf16}, with
plots of these wave functions given in Fig.1-2 (for P-quarkonia) and
Fig.3-4 (for V-quarkonia) in \cite{hluf16}. Thus we want to mention
that the complete 2-fermion wave functions,
$\Psi_{P,V}(\widehat{q})$ are ultimately expressed in terms of
summation over various Dirac structures multiplying a single scalar
function $\phi_{P,V}(\widehat{q})$, whose detailed algebraic
expressions (gaussian functions) are not approximations, but are
obtained as analytic solutions of the algebraic form of the mass
spectral equations for P and V quarkonia (with the complete spectrum
written down in terms of principal quantum number $N$ in Eq,(35-36)
of \cite{hluf16}) in an approximate harmonic oscillator basis.

The projected wave function for positive energy,
$\psi^{++}(\hat{q})$ is obtained by the operation of projection
operators on $\psi_{P}(\hat{q}')$ and $\psi_{V}(\widehat{q})$
respectively as in Eq. (8):
\begin{eqnarray}
&&\nonumber
\psi_{P}^{++}(\hat{q}')=\Lambda_{1}^{+}(\hat{q}')\frac{\slashed{P}'}{M'}\psi_{P}(\hat{q}')
 \frac{\slashed{P}'}{M'}\Lambda_{2}^{+}(\hat{q}')\\&&
 \psi_{V}^{++}(\hat{q})=\Lambda_{1}^{+}(\hat{q})\frac{\slashed{P}}{M}\psi_{V}(\hat{q})
 \frac{\slashed{P}}{M}\Lambda_{2}^{+}(\hat{q})
\end{eqnarray}
where
$\Lambda_{1,2}^{+}(\hat{q})=\frac{1}{2\omega}[\frac{\slashed{P}}{M}\omega\pm
 m\pm\slashed{\hat{q}})]$,  $\Lambda_{1,2}^{+}(\hat{q}')=\frac{1}{2\omega'}[\frac{\slashed{P}'}{M'}\omega'\pm
 m\pm\slashed{\hat{q}}')]$, are called as
projection operators and $\omega^{2} = m^{2} + \hat{q}^{2}$,
$\omega'^{2} = m^{2} + \hat{q}'^{2}$. The relativistic positive
energy wave function of equal mass pseudoscalar quarkonia in the
center of mass system can be express as:
\begin{equation}
\psi^{++}_{P}(\hat{q}')=\frac{N'_{P}\phi_{P}(\hat{q}')}{2}\gamma_{5}\left\{M'
\left[\frac{m^{2}}{\omega'^{2}}+\frac{m}{\omega'}
-\frac{\hat{q}'^{2}}{m\omega'}\right]-\slashed{P}'\left[\frac{m^{2}}{\omega'^{2}}+\frac{m}{\omega'}\right]+
\slashed{\hat{q}'}\slashed{P}'\left[\frac{m}{\omega'^{2}}+\frac{1}{\omega'}\right]\right\}
\end{equation}
And the relativistic positive energy wave function of equal mass
vector quarkonia, in the center of mass system can be written as:
\begin{eqnarray}
&&\nonumber\
\psi^{++}_{V}(\hat{q})=\frac{N_{V}\phi_{V}(\hat{q})}{2}\{M\left[\frac{m^{2}}{\omega^{2}}-
\frac{m}{\omega}+\frac{\hat{q}^{2}}{m\omega}
 \right]\slashed{\varepsilon}+M\hat{q}.\varepsilon\left[\frac{m}{\omega^{2}}-\frac{1}{\omega}\right]
 +\left[\frac{m^{2}}{\omega^{2}}-\frac{m}{\omega}\right]\slashed{\varepsilon}\slashed{P}\\&&
 +\hat{q}.\varepsilon\left[\frac{m}{\omega^{2}}-\frac{1}{\omega}\right]\slashed{P}
-\left[\frac{m}{\omega^{2}}-\frac{1}{\omega}\right]\slashed{P}\slashed{\varepsilon}\slashed{\hat{q}}+\frac{\hat{q}.\varepsilon}{m\omega}
 \slashed{\hat{q}}\slashed{P}-\frac{M\hat{q}.\varepsilon}{m\omega}\slashed{\hat{q}}\}
\end{eqnarray}
The conjugate of $\psi^{++}_{P}(\hat{q}')$ is evaluated from
$\overline{\psi}^{++}_{P}(\hat{q}')=\gamma^{0}(\psi^{++}_{P})^{+}\gamma^{0}$,
after a sequence of steps is expressed as:
\begin{equation}
\overline{\psi}^{++}_{P}(\hat{q}')=-\frac{N'_{P}\phi_{P}(\hat{q}')}{2}\gamma_{5}\left\{M'
\left[\frac{m^{2}}{\omega'^{2}}+\frac{m}{\omega'}
-\frac{\hat{q}'^{2}}{m\omega'}\right]-\slashed{P}'\left[\frac{m^{2}}{\omega'^{2}}+\frac{m}{\omega'}\right]+
\slashed{P}'\slashed{\hat{q}'}\left[\frac{m}{\omega'^{2}}+\frac{1}{\omega'}\right]\right\}
\end{equation}
Now, if we evaluate the total transition amplitude using equations
Eq. (12), Eq. (17) and Eq. (18), we get,
\begin{equation}
M_{fi}=-\frac{e_{q}N'_{P}N_{V}}{2M}
\int\frac{d^{3}\hat{q}}{(2\pi)^{3}}\phi_{P}(\hat{q}')\phi_{V}(\hat{q})\left[TR\right]
\end{equation}
where
\begin{eqnarray}
&&\nonumber\left[TR\right]=Tr\{-\gamma_{5}\slashed{P}\slashed{P}'
\left[\frac{m^{2}}{\omega'^{2}}+\frac{m}{\omega'}\right]\left[M\left[\frac{m^{2}}{\omega^{2}}-
\frac{m}{\omega}+\frac{\hat{q}^{2}}{m\omega}
 \right]\slashed{\epsilon}\slashed{\varepsilon}-
\left[\frac{m}{\omega^{2}}-\frac{1}{\omega}\right]\slashed{\epsilon}\slashed{P}\slashed{\varepsilon}\slashed{\hat{q}}
-\frac{M\hat{q}.\varepsilon}{m\omega}\slashed{\epsilon}\slashed{\hat{q}}\right]\\&&
+\gamma_{5}\slashed{P}\slashed{P}'\slashed{\hat{q}'}\left[\frac{m}{\omega'^{2}}+\frac{1}{\omega'}\right]\left[
M\hat{q}.\varepsilon\left[\frac{m}{\omega^{2}}-\frac{1}{\omega}\right]\slashed{\epsilon}
+\left[\frac{m^{2}}{\omega^{2}}-\frac{m}{\omega}\right]\slashed{\epsilon}\slashed{\varepsilon}\slashed{P}
+\frac{\hat{q}.\varepsilon}{m\omega}\slashed{\epsilon}\slashed{\hat{q}}\slashed{P}\right]\}
\end{eqnarray}
Evaluating trace over the gamma matrices, one can obtain the
expression:
\begin{equation}
\left[TR\right]=-4M\epsilon_{\mu\nu\alpha\beta}P_{\mu}P'_{\nu}\epsilon_{\alpha}\varepsilon_{\beta}
\left[\frac{m}{\omega'^{2}}+\frac{1}{\omega'}\right]
\left[\frac{m^{3}}{\omega^{2}}-\frac{m^{2}}{\omega}+(\frac{m}{\omega^{2}}-\frac{1}{\omega})\hat{q}.\hat{q}'\right]
\end{equation}
where, $P$ and $P'$ are momenta of the initial (V) and final (P)
quarkonia respectively. One can rewrite the transition amplitude
after evaluation of the gamma matrices as:
\begin{equation}
M_{fi}=2e_{q}N'_{P}N_{V}
\int\frac{d^{3}\hat{q}}{(2\pi)^{3}}\phi_{P}(\hat{q}')\phi_{V}(\hat{q})\left[\frac{m}{\omega'^{2}}+\frac{1}{\omega'}\right]
\left[\frac{m^{3}}{\omega^{2}}-\frac{m^{2}}{\omega}+(\frac{m}{\omega^{2}}-\frac{1}{\omega})\hat{q}.\hat{q}'\right]
\epsilon_{\mu\nu\alpha\beta}P_{\mu}P'_{\nu}\epsilon_{\alpha}\varepsilon_{\beta}
\end{equation}
The decay width for $V\rightarrow P \gamma$, can be expressed as
(see Ref.\cite{bhatnagar12}):
\begin{equation}
\Gamma_{V\rightarrow P\gamma}=\frac{|\vec{P}'|}{8\pi
M^{2}}\left|M_{fi}\right|^{2};\\
|\vec{P}'|=\frac{1}{2M}\left[M^{2}-M'^{2}\right]
 \end{equation}
 From Eq. (22), one can
obtain:
\begin{eqnarray}
&&\nonumber
  \left|M_{fi}\right|^{2}=\left|f(\hat{q})\right|^{2}
  (\epsilon_{\mu\nu\alpha\beta}P_{\mu}P'_{\nu}\epsilon_{\alpha}\varepsilon_{\beta})^{2}\\&&
  = (M^{4}+10M^{2}M'^{2}+M'^{4}) \left|f(\hat{q})\right|^{2}
\end{eqnarray}
where
\begin{equation}
f(\hat{q})=2e_{q}N'_{P}N_{V}\int\frac{d^{3}
  \hat{q}'}{(2\pi)^{3}}\phi_{P}(\hat{q}')\phi_{V}(\hat{q})\left[\frac{m}{\omega'^{2}}+\frac{1}{\omega'}
  \right]\left[\frac{m^{3}}{\omega^{2}}-\frac{m^{2}}{\omega}+(\frac{m}{\omega^{2}}-\frac{1}{\omega})\hat{q}.\hat{q}'\right]
  \end{equation}
One can obtain the expression of the decay rate of heavy quarkonia
for the process, $V\rightarrow P\gamma$, as:
\begin{equation}
  \Gamma_{V\rightarrow
  P\gamma}=\frac{1}{64\pi}\left[M^{3}+\frac{M'^{4}}{M}-M'^{2}M-\frac{M'^{6}}{M^{3}}\right]^{3}|f(\hat{q})|^{2}
\end{equation}
$M'$ and $M$ are the masses of the Pseudoscalar heavy quarkonia
($\eta_{c}$, $\eta_{b}$) and Vector heavy quarkonia ($J/\psi$,
$\Upsilon$) respectively. In the expression for $f(\hat{q})$,
$N_{V}$ and $N'_{P}$ are the BS normalizers for heavy vector and
pseudoscalar quarkonia, respectively, which are given in a simple
form as in Ref.\cite{hluf16}:
\begin{equation}
 N_{V} =\left[16mM_{V}\int\frac{d^{3}\hat{q}}{(2\pi)^{3}}
\frac{\hat{q}^{2}}{\omega^{3}}\phi_{V}^{2}(\hat{q})\right]^{-1/2}
\end{equation}
 and
\begin{equation}
N'_{P}=\left[\frac{16M_{P}}{m}\int\frac{d^{3}\hat{q}'}{(2\pi)^{3}}
\frac{\hat{q}'^{2}}{\omega'}\phi_{P}^{2}(\hat{q}')\right]^{-1/2}
\end{equation}
 The ground state (1S) wave function and the radial wave functions for the 2S
and 3S excitations for initial heavy vector quarkonia are written as
in Eq. (41) Ref. \cite{hluf16}:
\begin{eqnarray}
&&\nonumber
\phi_{V}(1S,\hat{q})=\frac{1}{\pi^{3/4}\beta_{V}^{3/2}}e^{-\frac{\hat{q}^{2}}{2\beta_{V}^{2}}}\\&&
\nonumber \phi_{V}(2S,\hat{q})=
(\frac{3}{2})^{1/2}\frac{1}{\pi^{3/4}\beta_{V}^{3/2}}(1-\frac{2\hat{q}^{2}}{3\beta_{V}^{2}})e^{-\frac{\hat{q}^{2}}{2\beta_{V}^{2}}}\\&&
\phi_{V}(3S,\hat{q})=(\frac{15}{8})^{1/2}\frac{1}{\pi^{3/4}\beta_{V}^{3/2}}
(1-\frac{20\hat{q}^{2}}{15\beta_{V}^{2}}+\frac{4\hat{q}^{4}}{15\beta_{V}^{4}})e^{-\frac{\hat{q}^{2}}{2\beta_{V}^{2}}}
\end{eqnarray}
and the ground state (1S) wave function and the radial wave
functions for the 2S and 3S excitations for final heavy pseudoscalar
quarkonia are written as in Eq. (41) Ref. \cite{hluf16}:
\begin{eqnarray}
&&\nonumber
\phi_{P}(1S,\hat{q}')=\frac{1}{\pi^{3/4}\beta_{P}^{3/2}}e^{-\frac{\hat{q}'^{2}}{2\beta_{P}^{2}}}\\&&
\nonumber \phi_{P}(2S,\hat{q}')=
(\frac{3}{2})^{1/2}\frac{1}{\pi^{3/4}\beta_{P}^{3/2}}(1-\frac{2\hat{q}'^{2}}{3\beta_{P}^{2}})e^{-\frac{\hat{q}'^{2}}
{2\beta_{P}^{2}}}\\&&
\phi_{P}(3S,\hat{q}')=(\frac{15}{8})^{1/2}\frac{1}{\pi^{3/4}\beta_{P}^{3/2}}
(1-\frac{20\hat{q}'^{2}}{15\beta_{P}^{2}}+\frac{4\hat{q}'^{4}}{15\beta_{P}^{4}})e^{-\frac{\hat{q}'^{2}}{2\beta_{P}^{2}}}
\end{eqnarray}
where $\hat{q}'=\hat{q}+\frac{1}{2}|\vec{P}'|$ and
$|\vec{P}'|=\frac{M_{V}}{2}(1-\frac{M^{2}_{P}}{M^{2}_{V}})$.\\

The inverse range parameters for pseudoscalar and vector meson
respectively are,
$\beta_{P}=(4\frac{m\omega^{2}_{q\bar{q}}}{\sqrt{1+2A_{0}(N+\frac{3}{2})}})^{\frac{1}{4}}$,
and
$\beta_{V}=(2\frac{m\omega^{2}_{q\bar{q}}}{\sqrt{1+2A_{0}(N+\frac{3}{2})}})^{\frac{1}{4}}$
and are dependent on the input kernel and contain the dynamical
information, and they differ from each other only due to spin-spin
interactions.

We had recently calculated the mass spectrum of ground and excited
states of P and V quarkonia in \cite{hluf16}.  The input parameters
employed in this calculation that were  fit from the mass spectrum
of ground state pseudoscalar and vector quarkonia in Ref.
\cite{hluf16} are given in table 1.
\begin{table}[h]
  \begin{center}
  \renewcommand{\tabcolsep}{12pt}
\begin{tabular}{llllll}
\hline
     $C_{0}$ &$\omega_{0}(GeV)$&$\Lambda(GeV)$&$A_{0}$&$m_{c}(GeV)$&$m_{b}(GeV)$\\ \hline
    0.210&0.150&0.200&0.010&1.490&5.070\\
    \hline
      \end{tabular}
      \end{center}
  \caption{Input parameters of BSE-CIA framework}
    \end{table}

The numerical values of inverse range parameters $\beta_{P}$, and
$\beta_{V}$ for various P and V quarkonia in the radiative
transitions studied in this paper are listed in the Table 2 below.

\begin{table}[h]
\begin{center}
\renewcommand{\tabcolsep}{25pt}
\begin{tabular}{llll}
  \hline
   $1^{--}$ state &$\beta_{V}$&$0^{-+}$ state &$\beta_{P}$\\\hline
    $J/\psi(1S)$&0.2466&$\eta_{c}(1S)$&0.3486\\
    $\psi(2S)$&0.2442&$\eta_{c}(2S)$&0.3454\\
    $\psi(3S)$&0.2420&$\eta_{c}(3S)$&0.3422\\
    $\Upsilon(1S)$&0.5066&$\eta_{b}(1S)$&0.7165\\
    $\Upsilon(2S)$&0.5018&$\eta_{b}(2S)$&0.7097\\
    $\Upsilon(3S)$&0.4972&$\eta_{b}(3S)$&0.7032\\ \hline
   \end{tabular}
   \end{center}
   \caption{$\beta_{P}$ and $\beta_{V}$ values for ground
state and excited states of $\eta_{c}$, $\eta_{b}$, $J/\psi$ and
$\Upsilon$ (in GeV units) in present calculation (BSE-CIA).}
     \end{table}

We had fixed the input parameters by studying the mass spectra for
P and V quarkonia for a number of states. However in Tables 3 and
4, below we list only the spectra of the quarkonia for which we
have done calculations of radiative decay widths in this paper.

\begin{table}[htbp]
\begin{center}
\begin{tabular}{lllllll}
  \hline
    &BSE - CIA &Expt.\cite{olive14}&Pot.
    Model\cite{bhagyesh11}&
   QCD sum rule\cite{veli12}&LQCD\cite{burch09}&Re.P.Model\cite{ebert13} \\\hline
    $M_{\eta_{c}(1S)}$& 2.9509 & 2.983$\pm$0.0007&2.980 & 3.11$\pm$0.52 &3.292&2.981 \\
    $M_{\eta_{c}(2S)}$& 3.7352 & 3.639$\pm$0.0013  &3.600& &4.240&3.635\\
    $M_{\eta_{c}(3S)}$& 4.4486 &   &4.060& &&3.989\\
    $M_{\eta_{b}(1S)}$& 9.0005 & 9.398 $\pm$0.0032
    &9.390 &9.66$\pm$ 1.65&7.377&9.398  \\
    $M_{\eta_{b}(2S)}$& 9.7215 & 9.999$\pm$0.0028   &9.947 & &8.202&9.990   \\
    $M_{\eta_{b}(3S)}$&10.4201 &   & 10.291& &&10.329\\
    \hline
     \end{tabular}
   \end{center}
   \caption{Masses of ground and radially excited
states of $\eta_c$ and $\eta_b$ (in GeV.) in present calculation
(BSE-CIA) along with experimental data, and their masses in other
models.}
\end{table}
\begin{table}[htbp]
  \begin{center}
\begin{tabular}{lllllll}
  \hline
    &BSE - CIA &Expt.\cite{olive14}&Rel. Pot. Model\cite{ebert13}&Pot. Model\cite{bhagyesh11}
    &BSE\cite{wang06}&LQCD\cite{kawanai15}\\\hline
    $M_{J/\psi(1S)}$&3.0974& 3.0969$\pm$ 0.000011&3.096       & 3.0969 &3.0969&3.099 \\
    $M_{\psi(2S)}$&3.6676& 3.6861$\pm$ 0.00034  &3.685& 3.6890&3.686 &3.653\\
    $M_{\psi(3S)}$&4.1945&4.03$\pm$ 0.001&4.039       & 4.1407&4.065 &4.099\\
    $M_{\Upsilon(1S)}$&9.6719&9.4603$\pm$ 0.00026 &9.460&9.4603 &9.460&\\
    $M_{\Upsilon(2S)}$&10.1926&10.0233$\pm$0.00031&10.023&9.9814 &10.029 &\\
    $M_{\Upsilon(3S)}$&10.6979&10.3552$\pm$0.00005&10.355&10.3195 &10.379 &\\
     \hline
     \end{tabular}
   \end{center}
   \caption{Masses of ground, radially and orbitally excited states of heavy vector quarkonium,
        $J/\psi$ and $\Upsilon$ in BSE-CIA along with their masses in other models and experimental
          data (all units are in GeV).}
     \end{table}

\textbf{Numerical results}\\ We use the same input parameters listed
in Table 1, to calculate the decay widths for the process
$V\rightarrow P +\gamma$. The results of radiative decay widths of
our model are listed in Table 5.
\begin{table}[htbp]
   \begin{center}
   \renewcommand{\tabcolsep}{16pt}
\begin{tabular}{lllll}
  \hline
      Transition&Our work&Expt.\cite{olive14}&LFM\cite{ho}&PM\cite{stan} \\\hline
    $\Gamma_{J/\psi(1S)\rightarrow\eta_{c}(1S)\gamma}$&2.0054&1.5687$\pm$0.011 &1.67$\pm$0.05&1.8 \\
    $\Gamma_{\psi(2S)\rightarrow\eta_{c}(2S)\gamma}$&0.5709&0.2093$\pm$0.002&&0.4\\
    $\Gamma_{\psi(3S)\rightarrow\eta_{c}(3S)\gamma}$&0.2984&&&\\
    $\Gamma_{\Upsilon(1S)\rightarrow\eta_{b}(1S)\gamma}$&0.3387& &$0.043^{+0.09}_{-0.03}$&0.001     \\
    $\Gamma_{\Upsilon(2S)\rightarrow\eta_{b}(2S)\gamma}$&0.1053& &&0.0002         \\
    $\Gamma_{\Upsilon(3S)\rightarrow\eta_{b}(3S)\gamma}$&0.0781&&&          \\
       \hline
     \end{tabular}
   \end{center}
   \caption{Radiative decay widths of equal mass heavy vector quarkonium of ground state (1S) and radially excited
states (2S, 3S) in present calculation (BSE-CIA) along with their
decay widths in other models and experimental data (all values are
in units of Kev).}
 \end{table}\\

\section{Discussions and conclusion}
We have employed a 3D reduction of BSE (with a $4\times 4$
representation for two-body ($q\overline{q}$) BS amplitude) under
Covariant Instantaneous Ansatz (CIA), and used it for calculating
the amplitudes and decay widths for ground and radially excited
states of vector ($J/\Psi$, and $\Upsilon$) quarkonia in the
process, $V\rightarrow P+\gamma$.

The numerical values of decay widths calculated in this BSE
framework for ($1S$, $2S$, $3S$) states of $J/\psi$ and $\Upsilon$
are shown in Table 5. The numerical calculation in this work has
been done using Mathematica. We first fit our parameters to the
ground state masses of $\eta_{c}$, $\eta_{b}$, $J/\psi$ and
$\Upsilon$. Using the input parameters (along with the input quark
masses) listed in Table 1, we obtained the best fit to these ground
state masses. The same set of parameters above was used to calculate
the masses of all the other (excited) states of $\eta_{c}$,
$\eta_{b}$, $J/\psi$ and $\Upsilon$, as well as the leptonic decay
constants of these states. Two-photon as well as two gluon decay
widths of $\eta_{c}$, and $\eta_{b}$ was further studied ( see our
recent paper \cite{hluf16}).

The results obtained for decay width of ground and radially excited
states of $J/\psi$ and $\Upsilon$ are somewhat on the higher side in
comparison to central values of data for these states, which might
be probably due to the absence of the negative energy part employed
in Eq.(11), where we have considered only the positive energy part
(as in Ref. \cite{wang3} employed earlier for heavy mesons).
However, from Table 5, a wide range of variation of radiative decay
widths of $J/\Psi(1S)$, and $\Psi(2S)$ states in different models
can be observed. However, the important thing is that our radiative
decay width values for vector quarkonia, show a marked decrease as
one goes from $1S$ to $3S$ state for $J/\psi$ and $\Upsilon$, which
is again in conformity with data and other models. We have also
given our predictions for radiative decay widths of $\Psi(3S)$,
$\Upsilon(1S)$, $\Upsilon(2S)$, and $\Upsilon(3S)$ states, for which
data is currently not yet available. The aim of doing this study was
to mainly test our analytic forms of wave functions in Eqs.(29-30)
obtained as solutions of mass spectral equations in an approximate
harmonic oscillator basis obtained analytically from the $4\times 4$
BSE as our starting point (that had so far given good predictions
for leptonic decays of P and V quarkonia, and the two-photon, and
the two-gluon decays of P-quarkonia \cite{hluf16}), to the single
photon radiative decays of V-quarkonia. This would in turn lead to
validation of our approach, which provides a much deeper insight
than the purely numerical calculations in $4\times 4$ BSE approach
that are prevalent in the literature.

In the process of arriving at analytic solution of spectra by
solving the coupled Salpeter equations, we have worked in
approximate harmonic oscillator basis, and also employing
approximation: $\omega\sim m$ for heavy quarks. We do concede that
some numerical accuracy has been lost in the process, but at the
same time we have obtained a much deeper understanding of the mass
spectra of quarkonia, where the equations are expressible in terms
of the principal quantum number $N$. We wish to mention that to the
best of our knowledge, we have not encountered any work in $4\times
4$ representation of BSE, that treats this problem analytically. On
the contrary all the other $4\times 4$ approaches adopt a purely
numerical approach of solving  the coupled set of Salpeter
equations, which may enhance the numerical accuracy, but this is at
the expense of a deeper understanding of the spectral problem. We
further wish to mention that the correctness of our approximations
can be judged by the fact that our plots of wave functions obtained
analytically for various states of P and V quarkonia \cite{hluf16}
are very similar to the corresponding plots of wave functions of
various states of these quarkonia obtained by purely numerical
approach in \cite{wang10}.

We are also not aware of any other BSE framework, involving
$4\times4$ BS amplitude, and with all the Dirac structures
incorporated (in fact many works use only the leading Dirac
structure), that treats these problems analytically, and uses the
algebraic forms of wave functions derived analytically from mass
spectral equation for calculation of various transitions. To the
best of our knowledge, all the other $4\times4$ BSE approaches treat
this problem numerically after obtaining the coupled set of
equations.


\end{document}